\documentstyle[twocolumn,aps,epsfig]{revtex}
\begin{document}
\draft
\twocolumn[\hsize\textwidth\columnwidth\hsize\csname @twocolumnfalse\endcsname
\title{The small polaron crossover transition in colossal 
magnetoresistance (CMR) manganites}
\author{Unjong Yu and B. I. Min}
\address{Department of Physics,
        Pohang University of Science and Technology, 
        Pohang 790-784, Korea}
\date{\today}
\maketitle

\begin{abstract}
Based on the combined model of the double exchange 
and the polaron, we have studied the small-to-large polaron 
crossover transition and 
explored its effects on the magnetic and transport
properties in colossal magnetoresistance (CMR) manganites.
We have used the variational Lang-Firsov canonical transformation, 
and shown that the magnetic and transport properties of both high and
low $T_C$ manganites are well described in terms of a single formalism. 
We have reproduced the rapid resistivity drop below $T_C$, 
a realistic CMR ratio, and the {\it first-order-like} 
sharp magnetic phase transition, which are observed in low $T_C$ manganites.
\end{abstract}

\pacs{75.30.Vn,71.38.+i,71.30.+h}
\vskip2pc]
\narrowtext

Despite  intensive researches on the colossal magnetoresistance (CMR)
manganites R$_{1-x}$A$_x$MnO$_3$ (R = rare-earth; A = divalent cation), 
there are many unusual properties that remain to be clarified.
The most outstanding feature of 
these compounds is the close relationship between transport
phenomena and the magnetism for $0.2 \lesssim x \lesssim 0.5$, 
which have been traditionally understood 
by the double exchange (DE) mechanism \cite{Zener,Anderson}. 
However, many recent experiments,  especially the giant isotope effect in
the magnetic transition temperature ($T_C$) \cite{Zhao},
as well as theoretical works \cite{Millis1} indicate that the 
electron-phonon interaction plays a crucial role in the physics of the 
CMR manganites. The small-to-large polaron transition near $T_C$
has been reported by various experimental tools
\cite{Billinge,Louca,Tyson,Booth,Lanzara}.

The R$_{0.7}$A$_{0.3}$MnO$_3$ CMR compounds have a wide range of $T_C$ from 70 K
to 365 K depending on the average radius of R and A cations
($\left\langle r_A \right\rangle$) \cite{Hwang}. Interestingly,
the low $T_C$ manganites are 
different from the high $T_C$ manganites in many properties.
The low $T_C$ manganites have higher resistivity,
much sharper resistivity peaks near $T_C$, and more enhanced 
magnetoresistance (MR). They often exhibit hysteresis behaviors 
in the temperature dependent resistivity \cite{Hwang}. 
Further, in contrast to the other 
manganites that experience a metal-insulator (M-I) transition near $T_C$,
the highest $T_C$ manganite La$_{0.7}$Sr$_{0.3}$MnO$_3$
is metallic in the whole temperature range.

%The \hfill magnetic \hfill moment \hfill measurement \hfill for \\
The magnetic moment measurement for 
La$_{0.67}$(Ba$_x$Ca$_{1-x}$)$_{0.33}$MnO$_3$ with varying $x$
shows that the magnetic transition becomes steeper for low $T_C$ samples,
reminiscent of the {\it first-order-like} phase transition 
for $x=0.0$ ($T_C$=340 K and 265 K for $x=1.0$ and $x=0.0$, respectively) 
\cite{Moutis}. 
Also the NMR measurement for Pr$_{0.7}$Ca$_{0.15}$Sr$_{0.15}$MnO$_3$ 
($T_C$=180 K) and Pr$_{0.7}$Ba$_{0.3}$MnO$_3$ (($T_C$=175 K) 
shows that the hyperfine fields on $^{55}$Mn
nuclei remain finite up to $T_C$, indicating the first order
magnetic transition \cite{Savosta}.
Neutron scattering experiment indicates that the spin dynamics near $T_C$
of low $T_C$ Nd$_{0.7}$Sr$_{0.3}$MnO$_3$ ($T_C$=197.9 K) 
is very distinct from that of high $T_C$ 
Pr$_{0.7}$Sr$_{0.3}$MnO$_3$ ($T_C$=300.9 K) \cite{Fernandez}. 
The spin wave stiffness data for the former show no evidence of the expected
spin wave collapse at $T_C$, as in La$_{0.67}$Ca$_{0.33}$MnO$_3$
\cite{Lynn}.  
It is pointed out that the exotic spin dynamics in the low $T_C$ manganite is 
related to the increased electron-lattice coupling \cite{Fernandez}.
 
Various theoretical models are proposed to explore the origin of the
CMR phenomena.  Among those, the combined model of the DE and the 
polaron explains the transport and other many experimental 
features rather well \cite{Roder,Millis2,Lee}. Indeed, this model could 
explain the resistivity and the magnetic behaviors of 
the high $T_C$ manganites such as La$_{0.7}$Ba$_{0.3}$MnO$_3$. 
However, it appears that this model cannot 
describe a rapid drop of the resistivity below $T_C$ and 
very large MR ratio of the low $T_C$ manganites 
such as La$_{0.7}$Ca$_{0.3}$MnO$_3$. 
Including the effect of the phonon hardening below $T_C$
improves the resistivity drop and the MR ratio substantially,
but the characteristic magnetic field needed to reproduce
the observed MR is still too high ($\sim 15-20$ T), as
compared to the experimental value ($\sim$ 4 T) \cite{Yu}.  
It is thus argued that the polaron model 
is not sufficient to explain the physics of CMR \cite{Alex}. 
Therefore the central issue is whether the other mechanism than
the the DE and the polaron should be invoked to understand the physics
of CMR manganites.
 
Motivated by the very distinct behaviors in the resistivity and 
the magnetism between the low and high $T_C$ manganites, 
we have reexamined the effects of the polaron transition 
in the combined model of the DE and the polaron.
Employing the variational Lang-Firsov (VLF) canonical 
transformation, we have investigated the 
small-to-large polaron transition
% in the small polaron model incorporating the DE
and explored its effects on magnetic and transport properties of 
both the low and high $T_C$ manganites. R\"{o}der {\it et al.} 
\cite{Roder} have tried a similar approach, but they did not show
the transport properties explicitly in connection with the CMR phenomena.
Millis {\it et al.} \cite{Millis2} have obtained comparable results 
using the dynamical mean field approximation, but the physics 
producing CMR is hidden and the resulting characteristic magnetic 
field needed to reproduce the observed MR is also too high.

In order to study the polaron transport incorporating the DE, we adopt the 
following Hamiltonian of the Zener-Holstein type:
\begin{eqnarray}
H &=& - t \gamma(T) \,
       \sum_{i \delta} c_{i+\delta}^{\dagger} c_i
  + \sum_{\bbox{q}} \omega_{q} a_{\bbox{q}}^{\dagger} a_{\bbox{q}} \nonumber \\
  &&+ \sum_{i \bbox{q}} c_{i}^{\dagger} c_i e^{i \bbox{q} \cdot \bbox{R}_i}
      M_{q} ( a_{\bbox{q}} + a_{-\bbox{q}}^{\dagger} ).
\label{hamiltonian}
\end{eqnarray}
Here the bandwidth variation by the DE is included in the hopping parameter 
with a factor $\gamma(T)=\left\langle\cos(\theta/2) \right\rangle$,
where $\theta$ is the angle between local spins of 
neighboring Mn sites. $M_{q}$ is the parameter representing the strength of
the electron-phonon interaction. 
For simplicity but without loss of the physics,
we consider the single $e_g$ orbitals and the single phonon modes, 
which is plausible to describe the metallic phase of 
CMR manganites \cite{Roder}.

In the strong coupling 
($E_p \equiv \sum_{q} {M_q^2}/{\omega_q} \gg t$)
and the non-adiabatic limit ($\omega_{q} \gg t$), one can apply
the conventional Lang-Firsov (LF) transformation and obtain the resistivity 
via the Kubo formula, considering the hopping term as a perturbation
\cite{Lee}.
In the opposite weak coupling limit, the analytic solution can be obtained
by treating the interaction term as a perturbation. It is, however,
difficult to evaluate physical quantities in the intermediate regime
inbetween.

We have adopted the VLF transformation to describe the small-to-large polaron 
transition and the associated physics in the intermediate regime. 
In the VLF method, one performs the Lang-Firsov transformation with some 
variational parameters and finds their values
from the minimization conditions of the total energy. 
We follow the method of Das and Sil \cite{Das} in which three parameters are
assumed:
on-site lattice distortion $\Delta$, nearest neighbor distortion $\Delta'$,
and two-phonon coherent state parameter $\alpha$. The coherent state
allows for the anharmonicity of the lattice fluctuations, which
become important at finite polaron densities and in 
the intermediate coupling and frequency regime \cite{Zheng}.  
The nearest neighbor 
distortion can be neglected in the three dimensional (3D) case for which the 
small-to-large polaron transition is so abrupt that $\Delta'$
does not contribute much in the course of the transition. 

Using the VLF transformation, we have
\begin{eqnarray}
\bar{H} &=& e^S H e^{-S}  \\
&=& - t \gamma(T) \,
    \sum_{j \delta} c_{j+\delta}^{\dagger} c_j X_{j+\delta}^{\dagger} X_j 
  + \sum_{\bbox{q}} \omega_{q} a_{\bbox{q}}^{\dagger} a_{\bbox{q}} 
     \nonumber \\ &&
  - E_p (2-\Delta)\Delta \sum_{j} n_j 
  + (1-\Delta) \sum_{j q} n_j M_q (a_{\bbox{q}} + a_{-\bbox{q}}^{\dagger}),
    \nonumber  
\end{eqnarray}
with
$
S = - \sum_{j \bbox{q}}  \frac{M_q}{\omega_q}  \Delta \, 
   c_j^{\dagger} c_j e^{i \bbox{q} \cdot \bbox{R}_j} 
   (a_{\bbox{q}} - a_{-{\bbox{q}}}^{\dagger}),$ and
$ 
X_j = \exp\!\left[ \sum_{\bbox{q}} \frac{M_q}{\omega_q}  \Delta \,
   e^{i {\bbox{q}} \cdot \bbox{R}_j} 
   (a_{\bbox{q}} - a_{-\bbox{q}}^{\dagger}) \right] .
$
Then, using the two-phonon coherent state 
$|\psi_{\rm ph}\rangle = e^{-\alpha \sum_{\bbox{q}} 
(a_{\bbox{q}} a_{-\bbox{q}} - a_{\bbox{q}}^{\dagger} a_{-\bbox{q}}^{\dagger})
} |0\rangle$ ($\alpha \geq 0)$  as a trial phonon wave function,
one can obtain the effective polaron Hamiltonian. Now the ground state 
energy is calculated in the framework of the mean-field theory:
\begin{eqnarray}
\bar{E}/N &=&
      - 2 z t \gamma(T) \, \exp\!\left[-\frac{E_p}{\omega_0} 
                                     \Gamma \right]
        n (1-n) 
      + \omega_0 \, \sinh^2 (2\alpha) 
       \nonumber \\ &&
      - n E_p (2- \Delta) \Delta ,  \label{energy}
\end{eqnarray}
where $\Gamma \equiv \Delta^2 e^{-4\alpha}$,
$z$ is the coordination number,
 and $n = N^{-1} \sum_i n_i$ is the carrier density.
We assumed a single phonon frequency $\omega_0$.
The values of $\Delta$ and $\alpha$ are obtained from the minimization of
Eq.\ (\ref{energy}).  
With $\Delta = 1.0$ and $\alpha = 0$, that is, with $\Gamma =1.0$,
one can reproduce the result of 
the conventional LF transformation. 
Note that, in the VLF formalism, the effective hopping parameter
is given by ${\tilde t}=t \gamma(T) \exp\!\left(-\frac{E_p}{\omega_0}
                                     \Gamma \right)$
with an additional parameter $\Gamma (\le 1.0)$ in the polaron narrowing 
factor, so that the narrowing effect becomes weakened in the VLF formalism. 

\begin{figure}
\centerline{\epsfig{figure=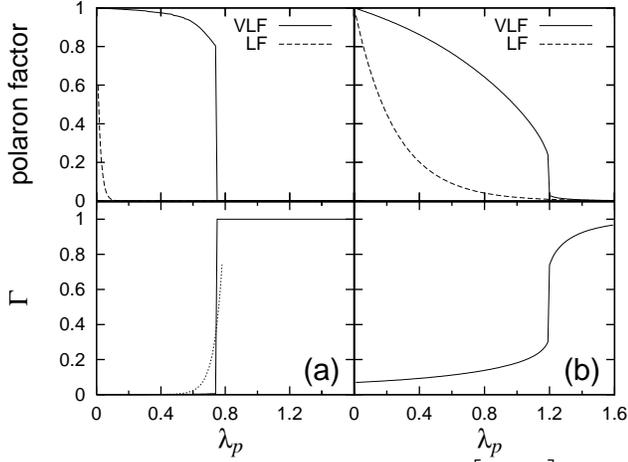,width=8.5cm}}
\caption{The polaron narrowing factor $\exp\!\left[-\frac{E_p}{\omega_0}
\Gamma \right]$
 and the parameter $\Gamma$ as a function of $\lambda_p$ for the 3D systems
 with $n=0.3$ obtained using the VLF transformation: 
(a) the adiabatic case ($\omega_0 / t = 0.13$) and (b) the non-adiabatic
case ($\omega_0 / t = 3.0$). Results of the conventional Lang-Firsov (LF)
 transformation ($\Gamma=1.0$) are also presented in the top panel
for comparison. Dots in the bottom panel of (a) represent an
illustration of the continuous well-behaved function smoothing the 
abrupt self-trapping transition.
 }
\label{vlf}
\end{figure}

Figure~\ref{vlf} provides $\Gamma$ and the polaron narrowing factor
as a function of electron-phonon coupling constant
$\lambda_p \equiv  E_p / (2 z t)$ for both the adiabatic and non-adiabatic
cases.  As shown in Fig.~\ref{vlf}, the large-to-small polaron 
transition, the so-called self-trapping transition takes place
near $\lambda_p \approx 1$ and is steeper for the adiabatic 
case. Also notable is that the critical value of $\lambda_p^c$ at which 
the transition occurs is smaller for the adiabatic case.
Though we do not show the results for 1D and 2D systems,  the transition 
for 3D systems is steeper.  These features
are consistent with those obtained using other methods 
such as the exact diagonalization \cite{Capone}, 
the dynamical mean field approximation \cite{Ciuchi}, and the
quantum Monte-Carlo method \cite{Raedt,Kornilovitch}. 
Since the CMR manganites are close to 3D adiabatic systems,
a very steep small-to-large polaron transition is expected.

It is tempting to interpret that the very steep self-trapping transition 
in Fig.~\ref{vlf} is the real first order phase transition.  
Caution is needed, however. According to the 
study of L\"owen \cite{Lowen} for the Hamiltonian 
of the Holstein type, the self-trapping transition should be
an analytical crossover not an abrupt (nonanalytical) phase transition.
Indeed, numerical calculations for various model systems
\cite{Capone,Ciuchi,Raedt,Kornilovitch,Fehske} show rather continuous
self-trapping transitions.
Then the seemingly nonanalytical transition seen in Fig.~\ref{vlf}
may not be real but result from the shortcomings of
the VLF method itself. 
Nevertheless, the general trend of the self-trapping transition
obtained in Fig.~\ref{vlf} can be
applied to CMR manganites for the analysis of transport and magnetic 
properties, by using a well-behaved function that smooth the
nonanalytical transition (dots in the bottom panel of Fig.~\ref{vlf}(a)).

To extend Eq.\ (\ref{energy}) to the finite temperature
regime, the entropy term originating
from the magnetic ordering must be included \cite{Kubo}.
The resulting free energy per site is represented by
\begin{eqnarray}
F &=&
      - 2 z t \gamma(\nu) \, \exp\!\left[-\frac{E_p}{\omega_0} 
                            \Gamma (2 N_0 +1) \right] n (1-n) 
       \nonumber \\ &&
      + \omega_0 \left\{ N_0 \left[ \sinh^2 (2\alpha) + \cosh^2 (2\alpha)
                 \right] + \sinh^2 (2\alpha) \right\}
       \nonumber \\ &&
      - n E_p (2- \Delta) \Delta 
      - k_B T \left\{ \ln [ Q_S(\nu) ] - 
                     \nu m_S (\nu) \right\}, \label{free}
\end{eqnarray}
where $\nu$ is the order parameter,
$Q_S(\nu) = \sum_{M = -S}^{S} \exp\!\left( \nu {M}/{S} \right)$ is
the partition function with spin $S=2$, and $m_S(\nu)$ is the normalized 
magnetization per site.  $N_0$ is the phonon distribution 
function and can be ignored in the temperature region of interest.
$\gamma(T)$ is also a function of $\nu$, being
constant (0.6) for $T>T_C$ and increasing below $T_C$ up to 0.8.
Now the free energy is to be minimized instead of Eq.\ (\ref{energy}).

\begin{figure}
\centerline{\epsfig{figure=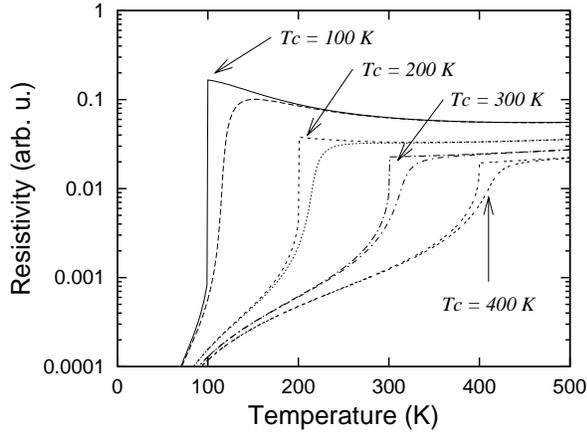,width=8.0cm}}
\caption{Magnetization vs. temperature for various $T_C$'s. 
%Temperature is normalized to $T_C$. 
Result in the DE only model is  also presented for comparison.
Parameters used in the calculation are 
  $t \gamma(T_C) = 0.3$ eV and $\omega_0 = 0.04$ eV.
 }
\label{magnetism}
\end{figure}

\begin{figure}
\centerline{\epsfig{figure=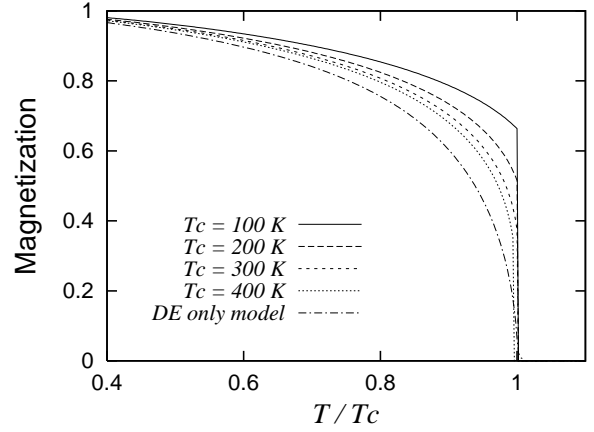,width=8.0cm}}
\caption{Temperature dependent resistivity without 
and with the magnetic field of 5 T (lower curves) for various $T_C$'s. 
Note that the $y$-axis is in the $\log$ scale.
Parameters used in the calculations are the same as in 
 Fig.~\protect\ref{magnetism}.
 }
\label{resistivity}
\end{figure}

Upon cooling below $T_C$, the DE 
hopping parameter $t \gamma(T)$ increases.  Accordingly, if
$\lambda_p$ decreases below $\lambda_p^c$, 
$\Gamma$ becomes reduced a lot to weaken the polaron narrowing effect 
as seen in Fig.~\ref{vlf}. This implies that
$\Gamma$ is an implicit function of the hopping parameter:
with a larger hopping having reduced $\Gamma$.
In consequence, the effective hopping parameter $\tilde t$
is rapidly enhanced to cause an enormous reduction of the resistivity 
below $T_C$ (see Fig~\ref{resistivity}).
In turn, this has an effect of increasing $\gamma(T)$  to
make the magnetic transition much steeper than in the case of
the DE only model.
This feature will be discussed more in Fig.~\ref{magnetism}.
The external magnetic field gives rise to the same effect, producing
the CMR along with the bandwidth increment (Fig.~\ref{resistivity}).

From Eq.\ (\ref{free}), $T_C$ for $S = 2$ can be derived:
\begin{eqnarray}
k_B T_C = \frac{9}{50} \: 2 z n (1-n) \;
     t \exp\!\left[ - \frac{E_p}{\omega_0} \, \Gamma \right].
 \label{tc}
\end{eqnarray}
This relation provides a possible 
account of the intriguing evolution of $T_C$ with
chemical pressure in R$_{0.7}$A$_{0.3}$MnO$_3$,
determined by $\left<r_A \right >$ \cite{Hwang}.
This behavior is qualitatively understood in terms
of the variation of the $\left<Mn-O\right>$ bond length with
$\left<r_A \right >$ and
the subsequent variation of the hopping parameter $t$ \cite{Rad}.
However, the observed variation in the bond 
length, at most 1\%, is thought to be too small to explain 
$\sim 500\%$ variation in $T_C$.
However, once employing the above polaron narrowing effects,
one can account for the huge variation of $T_C$ 
even with the small variation of the hopping parameter $t$.
That is, the increase of $\left<r_A \right >$ enhances $t$,
and so reduces $\lambda_p$. As a result, $\Gamma$
in the exponent of the polaron narrowing factor in Eq.\ (\ref{tc})
decreases so as to enhance $T_C$ a lot.
In  a similar way,
the decreasing isotope effect with $\left<r_A \right >$ can also be
understood by noting that the isotope exponent $\beta 
(\equiv -\Delta\ln T_C/ \Delta\ln M_O$, $M_O$: oxygen isotope mass)
is a decreasing function of the hopping parameter through $\Gamma$:
$\beta \sim \frac{E_p}{\omega_0} \Gamma(t)$ \cite{Zhao}.

Since we have determined the polaron narrowing factor for all
the range of $\lambda_p$ (Fig.~\ref{vlf}),
$\lambda_{p0}$ for given $T_C$ can be estimated using Eq.\ (\ref{tc}).
We have larger $\lambda_{p0}$ for lower $T_C$ to have a higher slope
of $\Gamma$ at $\lambda_{p0}$,  e.g., $\lambda_{p 0} = 0.646$ and 0.619
for $T_C =100$ K and 400 K, respectively, for parameters given in
Fig.~\ref{magnetism}. Hence, as $\gamma(T)$
increases below $T_C$, the subsequent variation of $\lambda_p$ 
drives a large reduction of $\Gamma$. Therefore one can expect 
a steeper magnetic transition and more rapid drop of the resistivity 
for low $T_C$ manganites.
Figure~\ref{magnetism} presents the temperature dependent
behaviors of the magnetization for various $T_C$. 
It is seen that the magnetic transition becomes steeper for
lower $T_C$. Thus, the magnetic transition 
for $T_C < 200$ K is almost like the 
first order phase transition. Note, by contrast, that the magnetic 
transition in the DE only model takes place much more smoothly reflecting
the second order phase transition. 
In view of these features, the change of magnetic properties in
La$_{0.67}$(Ba$_x$Ca$_{1-x}$)$_{0.33}$MnO$_3$ with varying $x$
can be elucidated \cite{Moutis}.  Further, this feature 
provides a clue to describe the exotic spin dynamics observed
in low $T_C$ manganites.
Since the transition is close to the first order phase transition,
the coexistence between the paramagnetic and the ferromagnetic phase
can occur in the course of the transition, manifesting 
features of the discontinuous drop of the hyperfine fields \cite{Savosta}
and the spin wave stiffness \cite{Fernandez} near $T_c$.

The routine evaluating the conductivity by means of the
Kubo formula is the same 
as in Ref.~\cite{Lee} except for $\Gamma$ in the polaron narrowing factor.
Figure~\ref{resistivity} shows the resulting resistivity as a function of 
temperature.  Here the phonon hardening effect below $T_C$
is also taken into account \cite{Yu}.
The M-I transition above $T_C$ is
clearly seen in the cases of low $T_C$.
For $T_C = 100$ K, the resistivity peak is very sharp and the MR is 
very large. Note that the $y$-axis is in the $\log$ scale so that the
resistivity drop below $T_C$ amounts to three orders of magnitude.
Hence the characteristic field for the observed MR becomes low enough,
in agreement with the experiment. In the case of $T_C=200$ K, we have
$\sim 94\%$ MR ratio for  H~=~5~T.
In addition, the lower $T_C$ manganites show higher resistivity 
in consistent with the experiments \cite{Hwang}. 
For higher $T_C$, the peak becomes dull, and eventually
the M-I transition disappears for $T_C$ larger than 300 K. 
This is again in agreement with the observation 
for La$_{0.7}$Sr$_{0.3}$MnO$_3$.  One must note that
the magnetic transition (Fig.~\ref{magnetism}) and the resistivity drop
(Fig.~\ref{resistivity}) are steeper for low $T_C$ manganites, 
demonstrating that both the CMR transport phenomena and the 
magnetic transition are closely correlated and 
are governed by the same physics.

Finally, it should be pointed out that the important effects, such as
the double degeneracy of the
$e_g$ orbitals, the inter-orbital Coulomb correlation,  and the
Jahn-Teller effects, are not considered in the present study.
One can in principle incorporating these effects, following the
formalism by Zang {\it et al} \cite{Zang}.
The Jahn-Teller interaction yields the effective DE
besides the Zener DE, and the Coulomb correlation gives rise to an
additional band narrowing effect. Therefore, these effects are expected to
enhance further the localization effect and the essential
results of the present study will not be altered.
The study that takes into account these effects 
explicitly is under progress.

In conclusion, we have investigated the small-to-large polaron crossover
transition for the combined model of the polaron and the DE interaction
and explored the effects of the polaron narrowing
on the magnetic and transport properties of both the low and 
high $T_C$ CMR manganites.  
We have shown that the polaron narrowing effect is much
more pronounced in the low $T_C$ manganites, which explains the rapid
resistivity drop, large MR ratio, and the {\it first-order-like}
sharp magnetic phase transition.

Acknowledgements$-$
Helpful discussions with J.D. Lee are greatly appreciated.
This work was supported by the KOSEF (96-0702-0101-3),
and in part by the BSRI program of the KME (BSRI-98-2438).

\newpage
\null
\newpage

\end{document}